\documentclass[epj]{svjour}

\begin{document}

\title{Spin-transport in multi-terminal normal metal - ferromagnet
systems with non-collinear magnetizations}

\author{Arne Brataas\inst{1} \and Yu.\ V. Nazarov\inst{2} \and Gerrit E.\ W.\ Bauer\inst{2} }

\offprints{Arne Brataas}

\institute{Harvard University, Lyman Laboratory of Physics, Cambridge,
MA 02138 \and Delft University of Technology, Laboratory of Applied
Physics and Delft Institute for Microelectronics, Lorentzweg 1, 2628
CJ Delft}

\abstract{
A theory of spin-transport in hybrid normal metal - ferromagnetic
electronic circuits is developed, taking into account non-collinear
spin-accumulation. Spin-transport through resistive elements is
described by 4 conductance parameters. Microscopic expression for
these conductances are derived in terms of scattering matrices and
explicitly calculated for simple models. The circuit theory is applied
to 2-terminal and 3-terminal devices attached to ferromagnetic
reservoirs.
}


\titlerunning{Spin-transport in multi-terminal normal
metal-ferromagnet systems}

\maketitle

\maketitle

\section{Introduction}

Spin-injection from a ferromagnetic metal into a non-magnetic metal
was first realized by Tedrow and Meservey in the
seventies\cite{Meservey94:173}. The discovery of the giant
magneto-resistance (GMR) in metallic magnetic multilayers has led to
an explosion of interest into spin-transport in hybrid normal
metal-ferromagnetic metal systems in the nineties (for reviews see
Ref.~\cite{Levy94:367}). Although discovered only ten years ago,
the GMR is commercially utilized in high-end magnetic recording
media. Spin-transport between ferromagnets through tunnel junctions
has also attracted renewed interest (for a review see
Ref.~\cite{Levy99:223}). Recently a magnetic double barrier
tunnel device was fabricated which enables the study of the interplay
between spin-polarized tunneling and Coulomb charging effects\cite
{Ono96:3449,Brataas99:93,Barnas98:1058}.

Magneto-electronic principles have until now led to magnetic-field
sensor devices. Future applications might include non-volatile memory
cells or even transistors, the latter being a three-terminal device.
Johnson realized that novel long-range effects in transport between
ferromagnets and normal metals can occur in multi-terminal
systems\cite{Johnson85:1790}. In Johnson's spin-transistor different
ferromagnetic (Ohmic) contacts provide information about the amount of
spin-accumulation in a normal metal film over distances much larger
than the mean-free path. A transistor-like effect was observed on
switching the configuration of the magnetizations of the ferromagnetic
contacts from parallel to anti-parallel.

Spin-transport in systems comprising ferromagnets with non-collinear
magnetization directions has attracted less attention. Moodera {\em et
al.}  measured the dependence of the current through magnetic tunnel
junctions on the relative angle between the magnetization directions
of the electrodes\cite{Moodera96:4724}. The results were in agreement
with the theoretical predictions based on the spin-torque model
introduced by Slonczewski\cite{Slon89:6995}. Non-collinear spin
transport has been addressed by a number of theoretical
papers\cite{Valet93:7099}.

The above examples do not exhaust the novel phenomena that can be
anticipated for multi-terminal hybrid normal metal-ferromagnet
circuits and devices. The question of the most appropriate theoretical
approach to the field arises. If the possibilities of macroscopic
quantum coherence and its application to quantum computing are
contemplated, a fully quantum mechanical treatment of the many-body
system is required, of course. However, for contact with the
experiments mentioned above and probably also with most to be realized
in the near future a full quantum mechanical treatment is unnecessary
and unrealistic. When at least a part of the device is diffusive
simplified approaches are called for. The situation is similar to that
in the field of inhomogeneous superconductors which has recently been
reviewed by Belzig {\em et al.}~\cite{Belzig99:1251}. The theoretical
framework of choice for transport in superconducting and/or magnetic
``dirty" systems is the non-equilibrium Keldysh Green functions
formalism in the quasiclassical approximation. However, it is
technically difficult and physically not very transparent for all but
the devoted specialist.  This led one of us to simplify equations for
complicated hybrid superconductor-normal metal systems to a handful of
easily accessible rules, the circuit theory of Andreev
reflection\cite{Nazarov94:1420}. We recently introduced a circuit (or
finite-element) theory of spin and charge transport in hybrid
ferromagnet-normal metal systems which provides parametric
dependencies of the electron transport properties as well as
microscopic expressions for the parameters and illustrated its
appeal\cite{Brataas00:2481}. Fig.~\ref{f:circuit} shows an example of
a typical many-terminal configuration of present interest.  The main
physical parameters of such a circuit are the magnetization
directions, the chemical potentials of the leads, and the contact
conductances between the normal and ferromagnetic metals. In this
paper we introduce the spin-polarized kinetic equations based on the
Keldysh Green function technique in the quasi-classical
approximation. When solved with judiciously chosen boundary conditions
the basic equations of the circuit theory emerge.

Before turning to the technical details let us first discuss the
conditions under which long-range spin effects are observable in
normal metals. Spins injected into a normal metal node relax due to
unavoidable spin-flip processes. Naturally the dwell time on the node
must be shorter than the spin-flip relaxation time in order to observe
non-locality in the electron transport. For a simple ferromagnet (F)
normal metal (N) double heterostructure (F|N|F) with anti parallel
magnetizations the condition can be quantified following
Ref.~\cite{Brataas99:93}. The spin-current into the normal metal
node is roughly proportional to the particle current,
$e(ds/dt)_{tr}\sim I=V/R$, where $s$ is the number of excess spins on
the normal metal node, $V$ is the voltage difference between the two
reservoirs coupled to the normal metal node, and $R$ is the F|N
contact resistance. When the node is smaller than the spin-diffusion
length, the spin-relaxation rate is
$e(ds/dt)_{rel}=-s/\tau_{\mbox{sf}}$, where $\tau_{\mbox{sf}}$ denotes
the spin-relaxation time on the node. (Otherwise this simple approach
breaks down since the spatial dependence of the spin-distribution in
the normal metal should be taken into account\cite{Huertas00:5700}).
The number of spins on the normal metal node is equivalent to a
non-equilibrium chemical potential difference $\Delta \mu = s \delta$
in terms of the energy level spacing $\delta$ (the inverse density of
states) (more generally the relation between $\Delta \mu$ and $s$ is
determined by the
spin-susceptibility\cite{Brataas99:93,MacDonald9912391}). The
spin-accumulation on the normal metal node significantly affects the
transport properties when the non-equilibrium chemical potential
difference is of the same order of magnitude or larger than the
applied source-drain voltage, $\Delta \mu >eV$ or $\delta
\tau_{\mbox{sf}}/h>R/R_{K}$, where $R_{K}=e^{2}/h$ is the quantum
resistance. We see that spin-accumulation is only relevant for
sufficiently small normal metal nodes and/or sufficiently long
spin-accumulation times and/or good contact conductances. The
implications of this relation have been discussed in
Ref.~\cite{Brataas99:93} for different materials.

In the present article we explain in more detail the foundations of
the circuit theory of spin-transport~\cite{Brataas00:2481} and present
more applications; a general result for two-terminal devices and
results for the spin-resistance in three-terminal devices similar to
the set-up of Johnson\cite{Johnson85:1790}. The manuscript is
organized in the following way. In Section~\ref{s:circuit} we describe
the basic entities in a circuit theory, the nodes, the contacts and
the reservoirs. The contact conductances are computed in
section~\ref{s:contact} for a diffusive, a ballistic and a tunnel
contact. The circuit theory is employed in
section~\ref{s:illustrations} to find the current trough two-terminal
systems and the 'spin-resistance' of a three terminal device. Our
conclusions can be found in
section~\ref{s:con}. Appendix~\ref{s:transform} gives the detailed
derivation of the transformation from a non-collinear to a collinear
two-terminal system.

\section{Circuit theory}
\label{s:circuit}

A typical (magneto)electronic circuit as schematically shown in
Fig.~\ref{f:circuit} can be divided into contacts (resistive
elements), nodes (low impedance interconnectors) and reservoirs
(voltage sources). The present theory is applicable when the contacts
limit the electric current and the nodes are characterized by a
distribution function which is constant in position and isotropic in
momentum space. The latter condition justifies a diffusion
approximation and requires that the nodes are either irregular in
shape or contain a sufficient number of randomly distributed
scatterers. State-of-the art magneto-electronic devices are rather
dirty, so this does not appear to be a major restriction for the
applicability of the theory. Because the spin-accumulation is not
necessarily parallel to the spin-quantization axis, the electron
distribution at each node is described by a $2 \times 2$ matrix in
spin-space. The current through each contact can be calculated as a
function of the distribution matrices on the adjacent nodes. The
spin-current conservation law then allows computation of the circuit
properties as a function of the applied voltages. The recipe for
calculating the current-voltage characteristics can be summarized as:
\begin{itemize}
\item Divide the circuit into nodes, contacts, and reservoirs.
\item Specify the $2\times 2$ distribution matrix
in spin-space for each node and reservoir.
\item Compute the current through a contact and the distribution
matrices in the adjacent nodes, which are related by the spin-charge
conductances specified below.
\item Make use of the spin-current conservation law at each node,
stating that the difference between in and outgoing spin-currents
equals the spin-relaxation rate.
\item Solve the resulting system of linear equations to obtain all
currents as a function of the reservoir chemical potentials.
\end{itemize}

\subsection{Node}

We denote the $2 \times 2$ distribution matrix at a given energy
$\epsilon$ in the node by $\hat{f}(\epsilon)$, where hat ($\hat{}$)
denotes a $2\times 2$ matrix in spin-space,. The external reservoirs
are assumed to be in local equilibrium so that the distribution matrix
is diagonal in spin-space and attains its local equilibrium value
$\hat{f}=\hat{1}f(\epsilon,\mu_{\alpha})$, $\hat{1}$ is the unit
matrix, $f(\epsilon,\mu_{\alpha})$ is the Fermi-Dirac distribution
function and $\mu_{\alpha}$ is the local chemical potential in
reservoir $\alpha$. The direction of the magnetization of the
ferromagnetic nodes is denoted by the unit vector ${\bf m}_{\alpha}$.

The $2 \times 2$ non-equilibrium distribution matrices in the nodes in
the stationary state are uniquely determined by current conservation
\begin{equation}
\sum_{\alpha} \hat{I}_{\alpha \beta} = \left( \frac{\partial
\hat{f}_{\beta}}{\partial t} \right)_{\mbox{rel}} \, , \label{curcons}
\end{equation}
where $\hat{I}_{\alpha \beta}$ denotes the $2 \times 2$ current in
spin-space from node (or reservoir) $\alpha$ to node (or reservoir)
$\beta$ and the term on the right hand side describes spin-relaxation
in the normal node. The right hand side of Eq.~(\ref{curcons}) can be
set to zero when the spin-current in the node is conserved, {\em i.e.}
when an electron spends much less time on the node than the spin-flip
relaxation time $\tau_{\mbox{sf}}$. If the size of the node in the
transport direction is smaller than the spin-flip diffusion length
$l_{\mbox{sf}} = \sqrt{D \tau_{\mbox{sf}}}$, where $D$ is the
diffusion coefficient then the spin-relaxation in the node can be
introduced as $(\partial \hat{f}^N / \partial t)_{\mbox{rel}} =
(\hat{1}\mbox{Tr}(\hat{f}^N)/2 - \hat{f}^N)/\tau_{\mbox{sf}}$.  If the
size of the node in the transport direction is larger than
$l_{\mbox{sf}}$ the simplest circuit theory fails and we have to use a
more complicated description with a spatially dependent
spin-distribution function\cite{Huertas00:5700}.

\subsection{Current through a contact}
\label{s:current}

A schematic picture of a contact between a normal metal and a
ferromagnetic node is shown in Fig.~\ref{f:contact}. The current is
evaluated on the normal side of the contact (dotted line).  The
current through the contact is
\begin{eqnarray}
\hat{I} & = & \frac{e}{h}\{ \sum_{nm}[ \hat{t}^{\prime
nm}\hat{f}^{F}\left( \hat{t}^{\prime mn}\right)^{\dagger} \nonumber \\
&& -(M\hat{f}^{N} - \hat{r}^{nm}\hat{f} ^{N}\left(
\hat{r}^{mn}\right)^{\dagger })] \} \, ,
\label{curLB}
\end{eqnarray}
where $r_{ss^{\prime }}^{nm}$ is the reflection coefficient for
electrons from transverse mode $m$ with spin $s^{\prime }$ incoming
from the normal metal side reflected to transverse mode $n$ with spin
$s$ on the normal metal side, and $t_{ss^{\prime }}^{\prime nm}$ is
the transmission coefficient for electrons from transverse mode $m$
with spin $s^{\prime}$ incoming from the ferromagnet transmitted to
transverse mode $n$ with spin $s$ on the normal metal side. (Note that
the hermitian conjugate in (\ref{curLB}) operates in the spin-space
and the space spanned by the transverse modes, {\em e.g.}
$(\hat{r}^{mn})^{\dag}_{ss'}=(\hat{r}^{nm}_{s's})^{\ast}$) The
relation (\ref{curLB}) can be found intuitively in the spirit of the
Landauer-B\"{u}ttiker formalism,\cite{Buttiker86:1761} but can also be
derived more rigorously by using the Keldysh formalism for
non-equilibrium transport. We use below the latter approach and
clarify the order of the spin-indices. Implicitly included in
(\ref{curLB}) are also effects related to the precession of spins
non-collinear to the magnetization direction in ferromagnets which
will be made more explicit below. 

The relation (\ref{curLB}) between the current and the distributions
has a simple form after transforming the spin-quantization axis. The
detailed calculation of this transformation is shown in
Appendix~\ref{s:transform}. Disregarding spin-flip processes in the
contacts, the reflection matrix for an incoming electron from the
normal metal transforms as
$$
\hat{r}^{nm}=\sum_s\hat{u}^s r_s^{nm} \, ,
$$
where $r_{\uparrow}^{nm}$ ($s=\uparrow$) and $r_{\downarrow}^{nm}$
($s=\downarrow$) are the spin-dependent reflection coefficients in the
basis where the spin-quantization axis is parallel to the
magnetization in the ferromagnet, the spin-projection matrices are
\begin{equation}
\hat{u}^{\uparrow} =  (\hat{1} + \hat{\mbox{\boldmath $\sigma$}}
\cdot {\bf m})/2 \, ,
\label{spinproj_up}
\end{equation}
\begin{equation}
\hat{u}^{\downarrow}  =  (\hat{1} - \hat{\mbox{\boldmath $\sigma$}}
\cdot {\bf m})/2 
\label{spinproj_do}
\end{equation}
and $\hat{\mbox{\boldmath $\sigma$}}$ is a vector of Pauli matrices.
Similarly for the transmission matrix
$$
\hat{t}^{\prime nm} (\hat{t}^{\prime mn})^{\dag}= \sum_{s} \hat{u}^s
|t^{\prime nm}_{s}|^2 \, ,
$$
where $t_{\uparrow}^{nm}$ and $t_{\downarrow}^{nm}$ are the
spin-dependent transmission coefficients in the basis where the
spin-quantization axis is parallel to the magnetization in the
ferromagnet.  Using the unitarity of the scattering matrix, we find
that the general form of the relation (\ref{curLB}) reads
\begin{eqnarray}
e\hat{I} & = &
G^{\uparrow} \hat{u}^{\uparrow} \left( \hat{f}^F - \hat{f}^N \right)
\hat{u}^{\uparrow}
+ G^{\downarrow} \hat{u}^{\downarrow} \left( \hat{f}^F - \hat{f}^N
\right) \hat{u}^{\downarrow} \nonumber \\
&& - G^{\uparrow \downarrow} \hat{u}^{\uparrow} \hat{f}^N
\hat{u}^{\downarrow}
- (G^{\uparrow \downarrow})^{\ast} \hat{u}^{\downarrow} \hat{f}^N
\hat{u}^{\uparrow} \, ,
\label{curgen}
\end{eqnarray}
where we have introduced the spin-dependent conductances
$G^{\uparrow}$ and $G^{\downarrow}$
\begin{eqnarray}
G^{\uparrow} = \frac{e^2}{h} \left[ M-\sum_{nm} |r^{nm}_{\uparrow}|^2
\right] = \frac{e^2}{h} \sum_{nm} |t^{nm}_{\uparrow}|^2 \label{Gup} \,
,
\end{eqnarray}
\begin{eqnarray}
G^{\downarrow} = \frac{e^2}{h} \left[ M-\sum_{nm}
|r^{nm}_{\downarrow}|^2 \right] = \frac{e^2}{h} \sum_{nm}
|t^{nm}_{\downarrow}|^2 \label{Gdo}
\end{eqnarray}
and the mixing conductance
\begin{equation}
G^{\uparrow \downarrow}=\frac{e^2}{h} \left[ M-\sum_{nm}
r_{\uparrow}^{nm} (r_{\downarrow}^{nm})^{\ast} \right] \, .
\label{Gmixing}
\end{equation}
The precession of spins leads to an effective relaxation of spins
non-collinear to the local magnetization in ferromagnets and
consequently the distribution function is limited to the form
$\hat{f}^F=\hat{1} f_0^F+\hat{\mbox{\boldmath $\sigma$}} \cdot {\bf m}
f_s^F$. Such a restriction does not appear in the normal metal node
and $\hat{f}^N$ can be any hermitian $2 \times 2$ matrix.

We thus see how the relation between the current through a contact and
the distributions in the ferromagnetic node and the normal metal node
are determined by 4 parameters, the two real spin-dependent
conductances ($G^{\uparrow}$, $G^{\downarrow}$) and the real and
imaginary parts of the mixing conductance $G^{\uparrow
\downarrow}$. These contact-specific parameters can be obtained by
microscopic theory or from experiments. The spin-conductances
$G^{\uparrow}$ and $G^{\downarrow}$ have been used in descriptions of
spin-transport for a long time\cite{Levy94:367}. The {\em mixing
conductance} is a new concept which is relevant for transport between
non-collinear ferromagnets. Note that although the mixing conductance
is a complex number the $2\times 2$ current in spin-space is hermitian
and consequently the current and the spin-current in any direction
given by Eq.~(\ref{curgen}) are real numbers. From the definitions of
the spin-dependent conductances (\ref{Gup}), (\ref{Gdo}) and the
`mixing' conductance (\ref{Gmixing}) we find
$$
2\mbox{Re} G^{\uparrow \downarrow} = G^{\uparrow} + G^{\downarrow} +
\frac{e^2}{h} \sum_{nm} |r_{\uparrow}^{nm}-r_{\downarrow}^{nm}|^2
$$
and consequently the conductances should satisfy
\begin{equation}
2\mbox{Re} G^{\uparrow \downarrow} \ge G^{\uparrow}+G^{\downarrow} \,
.
\label{rel}
\end{equation}
A physical interpretation of this result is given below.

Before outlining the derivation of (\ref{curLB}) leading to
(\ref{curgen}), let us discuss the physics in simple terms. Some
insight can be gained by re-writing the current and the distribution
function in terms of a scalar particle and a vector spin-contribution,
$\hat{I}=(\hat{1} I_0 +\hat{\mbox{\boldmath $\sigma$}} \cdot {\bf
I}_s)/2$, $\hat{f}^N=\hat{1} f_0^N +\hat{\mbox{\boldmath $\sigma$}}
\cdot {\bf s} f^N_s$ and $\hat{f}^F=\hat{1} f_0^F+\hat{\mbox{\boldmath
$\sigma$}} \cdot {\bf m} f^F_s$. The particle current can then be
written as
\begin{eqnarray}
I_0 & = & (G^{\uparrow}+G^{\downarrow}) (f_0^F-f_0^N) \nonumber \\
&& + (G^{\uparrow}-G^{\downarrow}) (f^F_s-{\bf m} \cdot {\bf
s} f^N_s) .  
\end{eqnarray}
The familiar expressions for collinear transport are recovered when
${\bf m} \cdot {\bf s}= \pm 1$. The spin-current is
\begin{eqnarray}
{\bf I}_s & = & {\bf m} [ (G^{\uparrow}-G^{\downarrow}) (f_0^F-f_0^N)
\nonumber \\
&& + (G^{\uparrow}+G^{\downarrow}) f^F_s +(2\mbox{Re} G^{\uparrow
\downarrow} - G^{\uparrow} - G^{\downarrow}) {\bf s} \cdot {\bf m}
f^N_s] \nonumber \\
&& -{\bf s} 2 \mbox{Re} G^{\uparrow\downarrow} f^N_s + ({\bf s} \times
{\bf m}) 2 \mbox{Im} G^{\uparrow\downarrow} f^N_s \, .
\label{spincurrent}
\end{eqnarray}
The first three terms point in the direction of the magnetization of
the ferromagnet ${\bf m}$, the fourth term is in the direction of the
non-equilibrium spin-distribution ${\bf s}$, and the last term is
perpendicular to both ${\bf s}$ and ${\bf m}$.  The last contribution
solely depends on the imaginary part of the mixing conductance. We can
interpret this term by considering how the direction of the spin on
the normal metal node ${\bf s}$ would change in time keeping, all
other parameters constant. The cross product creates a precession of
${\bf s}$ around the magnetization direction ${\bf m}$ of the
ferromagnet similar to a classical torque while keeping the magnitude
of the spin-accumulation constant. In contrast, the first four terms
represent diffusion-like processes which decrease the magnitude of the
spin-accumulation. We now understand condition (\ref{rel}) since
(\ref{spincurrent}) implies that the non-equilibrium spin-distribution
$f^N_s$ propagates easier into a configuration parallel to ${\bf s}$
than parallel to ${\bf m}$, since these processes are governed by {\em
positive} diffusion-like constants $2\mbox{Re} G^{\uparrow
\downarrow}$ and $2\mbox{Re} G^{\uparrow \downarrow} - G^{\uparrow} -
G^{\downarrow}$, respectively.

\subsection{Derivation of the current}

In this section the relation between the current through a contact and
the adjacent distribution functions (\ref{curLB}) is derived. This
derivation is not crucial for the understanding of Sections
\ref{s:contact}-\ref{s:con} and can be skipped by readers mainly
interested in the physical implications of (\ref{curLB}) and
(\ref{curgen}). We follow the lines of the derivation of the contact
current between a normal metal and a superconductor in
Ref.~\cite{Nazarov94:1420}. We consider a F-N system comprising
of a ferromagnet with arbitrary magnetization direction and a normal
metal node separated by a contact, as shown in Fig.~\ref{f:contact}.
The Stoner Hamiltonian is
\begin{equation}
\hat{H}=\left[ -\frac{1}{2m}\nabla ^{2}+V^{\mbox{p}}({\bf r})\right]
\hat{1}+\hat{V}^{\mbox{s}}({\bf r}),
\label{ham}
\end{equation}
where $V^{\mbox{p}}({\bf r})$ is the spin-independent potential and
$\hat{V}^{ \mbox{s}}({\bf r})$ is the spin-dependent potential. The
latter vanishes in the normal metal node and has an arbitrary, but
spatially independent direction in spin-space in the ferromagnetic
reservoir, $\hat{V}^{\mbox{s}}({\bf r})=(\mbox{\boldmath $\sigma$}
\cdot {\bf m}) V^{s}({\bf r})$ . Non-equilibrium transport properties
are most conveniently discussed in the framework of the Keldysh
formalism. The Keldysh Green function is given by the ($4\times 4$)
matrix
$$
\check{G}=\left( 
\begin{array}{cc}
\hat{G}^{R} & \hat{G}^{K} \\
\hat{0} & \hat{G}^{A}
\end{array}
\right) ,
$$
where $\hat{G}^{R}$, $\hat{G}^{K}$ and $\hat{G}^{A}$ are the retarded,
Keldysh and advanced Green functions respectively, which are ($2\times
2$) matrices in spin-space and $\widehat{0}$ is the ($2\times 2$) zero
matrix.  The retarded Green function in spin-space is
$$
\hat{G}^{R}(1,1^{\prime })=\left(
\begin{array}{cc}
G_{\uparrow \uparrow }^{R}(1,1^{\prime }) & G_{\uparrow \downarrow
}^{R}(1,1^{\prime }) \\
G_{\downarrow \uparrow }^{R}(1,1^{\prime }) & G_{\downarrow \downarrow
}^{R}(1,1^{\prime })
\end{array}
\right)
$$
and there are analogous expressions for $\hat{G}^{K}$ and
$\hat{G}^{A}$.  Here $1$ denotes the spatial and the time coordinates,
$1={\bf r}_{1}t_{1}$.  The symbol ``check'' ($\check{}$) denotes
($4\times 4$) matrices in Keldysh space and the symbol ``hat''
($\hat{}$) denotes ($2\times 2$) matrices in spin-space. The
spin-components of the Green functions are
\begin{eqnarray}
G_{\sigma s}^{R}(1,1^{\prime }) &=&-i\theta (t_{1}-t_{1^{\prime
}})\left\langle \left[ \psi _{\sigma }(1),\psi _{s}^{{\dag }}(1^{\prime })%
\right] \right\rangle_{+} , \\
G_{\sigma s}^{A}(1,1^{\prime }) &=&i\theta (t_{1^{\prime
}}-t_{1})\left\langle \left[ \psi _{\sigma }(1),\psi _{s}^{{\dag }%
}(1^{\prime })\right] \right\rangle_{+} , \\
G_{\sigma s}^{K}(1,1^{\prime }) &=&-i\left\langle \left[ \psi _{\sigma
}(1),\psi _{s}^{{\dag }}(1^{\prime })\right] \right\rangle \, ,
\end{eqnarray}
where $\psi_{s}^{\dag}(1)$ is the electron field operator for an
electron with spin $s$ in the $z$-direction.  The Keldysh Green
functions is determined by the equation
$$
\left( \hat{H}-i\hbar \frac{\partial }{\partial t}\right)
\check{G}({\bf r}t, {\bf r}^{\prime }t^{\prime })=\check{1}\delta
({\bf r}t-{\bf r}^{\prime }t^{\prime }),
$$
and the boundary conditions to be discussed below ($\check{1}$ is a $4
\times 4$ unit matrix). In the stationary situation
$\check{G}(1,1^{\prime })=\int d(E/2\pi )\exp (iE\left(
t_{1}-t_{1^{\prime }}\right) )\check{G}_{E}(r,r^{\prime })$ and the
Green function on a given energy shell is determined from 
$$
\left(
\hat{H}-E\right) \check{G}_{E}({\bf r},{\bf r}^{\prime })=\check{1}
\delta ({\bf r}-{\bf r}^{\prime })  \,  .
$$
We will in the following omit the index $E$ in denoting the Green
function at a given energy. The Keldysh Green function can be
decomposed into quasi-one-dimensional modes as
\begin{eqnarray}
\check{G}_{ss^{\prime }}({\bf r},{\bf r}^{\prime }) & &
=\sum_{nm,\alpha \beta } \tilde{G}_{nsms^{\prime }}^{\alpha \beta
}(x,x^{\prime }) \times \nonumber \\
&& \chi^n_{s}({\bf \rho };x{\bf )\chi }^{m \ast}_{s^{\prime }}({\bf
\rho }^{\prime };x^{\prime }) e^{i\alpha k_{s}^{n}x-i\beta
k_{s^{\prime}}^{m}x^{\prime }} \, ,
\end{eqnarray}
where $\chi^{n}_{s}({\bf \rho };x{\bf )}$ is the transverse wave
function and $k^{n}_{s}$ denotes the longitudinal wave-vector for an
electron in transverse mode $m$ with spin $s$. The indices $\alpha$
and $\beta$ denote right-going ($+)$ and left-going ($-$) modes. The
symbol `tilde' ($\tilde{ }$) denotes matrices in Keldysh space,
spin-space, ant the space spanned by the transverse modes and the
directions of propagation.

The current operator can be found from the continuity relation for the
electron density. The spin-density matrix is
$$
\rho _{\sigma s}(1)=\left\langle \psi _{s}^{{\dag }}(1)\psi _{\sigma
}(1)\right\rangle \, .
$$
The time-evolution of the spin-density matrix reads
$$
\frac{\partial }{\partial t_{1}}\rho _{\sigma s} = -\frac{\partial
}{\partial {\bf r}_{1}}{\bf J}_{\sigma s}^{\mbox{p} }+\left(
\frac{\partial \rho _{\sigma s}}{\partial t_{1}}\right)_{
\mbox{prec.}}  \, ,
$$
where we have inserted the Hamiltonian (\ref{ham}) and found the
spin-current
$$
{\bf J}_{\sigma s}^{\mbox{p}}=\frac{\hbar i}{2m}\left\langle \frac{
\partial \psi _{s}^{{\dag }}}{\partial {\bf r}_{1}}\psi _{\sigma
}-\psi _{s}^{{\dag }}\frac{\partial \psi _{\sigma }}{\partial
{\bf r }_{1}}\right\rangle ,
$$
and the spin-precession
\begin{equation}
\left( \frac{\partial \rho_{\sigma s}}{\partial t_{1}}\right)
_{\mbox{prec.}}=\frac{1}{i\hbar }\sum_{\alpha }\left[
V^{\mbox{s}}_{\sigma \alpha }\rho _{\alpha s}-\rho _{\sigma
\alpha }V^{\mbox{s}}_{\alpha s}\right] \, .
\label{spinprecession} 
\end{equation}
In three dimensions the spin-precession is an average over many states
with different Larmor frequencies which average out very quickly in a
ferromagnet, leading to an efficient relaxation of the non-diagonal
terms in the spin-density matrix that represent a spin-accumulation
non-collinear to the magnetization in the ferromagnet.  This
spin-relaxation mechanism does not exist in normal metals where in the
absence of spin-flip scattering the spin-wave functions remain
coherent.

In the stationary situation the spin-current is
\begin{eqnarray}
{\bf J}_{\sigma s} && =\left( \frac{\partial }{\partial {\bf r}_{1}}-
\frac{\partial }{\partial {\bf r}_{1^{\prime }}}\right) \frac{\hbar
i}{2m} \int \frac{dE}{2\pi }\int d(t_{1}-t_{1^{\prime }}) \times
\nonumber \\
&& \exp(iE(t_{1}-t_{1^{\prime }}))\left\langle \psi _{s}^{{\dag
}}(1)\psi _{\sigma }(1^{\prime })\right\rangle |_{{\bf r}_{1^{\prime
}}={\bf r}_{1}} \, .
\end{eqnarray}
We now define the extended $4 \times 4$ current matrix in Keldysh and
spin-space as
$$
\check{I}(x)=\int d{\bf \rho }\frac{e\hbar }{m}\left( \frac{\partial
}{ \partial x}-\frac{\partial }{\partial x^{\prime }}\right)
\check{G}({\bf r},{\bf r}^{\prime })|_{{\bf r}^{\prime }={\bf r}}.
$$
The transverse wave function $\chi^{n}_{s}({\bf \rho };x{\bf )}$ is
spatially independent in the leads and we find
\begin{eqnarray}
\check{I}_{ss^{\prime }}(x) && =ie\sum_{n\alpha \beta }\left( \alpha
v_{s}^n-\beta v^m_{s^{\prime }}\right) \tilde{G}_{nsms^{\prime
}}^{\alpha \beta }(x,x) \times \nonumber\\
&& \int d{\bf \rho }\chi^n_{s}({\bf \rho };x{\bf )\chi }^{m
\ast}_{s^{\prime }}({\bf \rho };x) \, ,
\end{eqnarray}
where $v^n_s=\hbar k^n_s/m$ is the longitudinal velocity for an
electron in transverse mode $n$ with spin $s$. In a normal metal, the
transverse states and the longitudinal momentum are spin-independent
and the Keldysh current simplifies to
\begin{equation}
\check{I}_{ss^{\prime}}(x)=2ie\sum_{n\alpha }\alpha
v^{n}\tilde{G}_{nsns^{\prime}}^{\alpha \alpha }(x,x) \, ,
\label{normcurr}
\end{equation}
which we will use to calculate the spin-current on the normal side of
the contact. We use the representation
\begin{eqnarray}
i\tilde{G}_{nsms^{\prime }}^{\alpha \beta }(x,x^{\prime })& =
&\frac{\tilde{g} _{nsms^{\prime }}^{\alpha \beta }(x,x^{\prime
})}{\sqrt{v^n_{s}v^m_{s^{\prime }}}} \nonumber \\
&& +\check{1}\delta _{ss^{\prime }}\frac{\alpha \delta _{\alpha ,\beta
} \mbox{sign}(x-x\prime )}{v^n_{s}} \, ,
\end{eqnarray}
where the latter term does not contribute to the current on the normal side,
and we have on the normal metal side
\begin{equation}
\check{I}_{ss^{\prime}}(x)=2e\sum_{n\alpha }\alpha
\tilde{g}_{nsns^{\prime}}^{\alpha \alpha }(x,x) \, .
\label{current}
\end{equation}

We now introduce the transfer matrix $M$ between waves propagating to
the right (left) on the right hand side of the contact $\Psi_R^+$
($\Psi_R^-$) and waves propagating to the right (left) on the left
hand side of the contact $\Psi_L^+$ ($\Psi_L^-$)
$$
\left( 
\begin{array}{c}
\Psi _{R}^{+} \\ 
\Psi _{R}^{-}
\end{array}
\right) 
=
M
\left( 
\begin{array}{c}
\Psi _{L}^{+} \\ 
\Psi _{L}^{-}
\end{array}
\right) \, .
$$
The elements of the transfer matrix are related to the reflection and
transmission coefficients by
$$
\left(
\begin{array}{cc}
m^{++} & m^{+-} \\
m^{-+} & m^{--}
\end{array}
\right)
=
\left(
\begin{array}{cc}
t-r'(t')^{-1} r & r' (t')^{-1} \\
-(t')^{-1}r & (t')^{-1}
\end{array} 
\right) \, ,
$$
Similarly we can introduce the scattering matrix
$$
S=
\left(
\begin{array}{cc}
r & t' \\
t & r'
\end{array} 
\right) \nonumber
$$
so that 
$$
S
=
\left(
\begin{array}{cc}
-(m^{--})^{-1} m^{-+} & (m^{--})^{-1} \\
m^{++} - m^{+-} (m^{--})^{-1} m^{-+} & m^{+-} (m^{--})^{-1}
\end{array}
\right)  \, ,
$$
where $r_{nm}^{s\sigma }$ is the reflection matrix for incoming states
from the left in mode $m$ and spin $\sigma $ to mode $n$ with spin
$s$, $t_{nm}^{s\sigma }$ is the transmission matrix for incoming
states from the left transmitted to outgoing states to the right,
$r^{\prime }$ is the reflection matrix for incoming states from the
right reflected to the right, and $t^{\prime }$ is the transmission
matrix for incoming states from the right transmitted to the
left. Unitarity of the $S$-matrix requires $S^{\dag} S=1$ and $S
S^{\dag}=$ and implies that the transfer matrix satisfies
\begin{eqnarray}
\bar{M}^{\dagger }\bar{\Sigma}_{z}\bar{M} & = & \bar{\Sigma}_{z} \, , \\
\bar{M}\bar{\Sigma}_{z}\bar{M}^{\dagger } & = & \bar{\Sigma} _{z} \, .
\end{eqnarray}
where $\left( \bar{\Sigma}_{z}\right) _{nsms^{\prime}}^{\alpha \beta
}=\alpha \delta _{\alpha ,\beta }\delta _{ns,ms^{\prime}}$ is a Pauli
matrix with respect to the direction of propagation.  In order to
connect the Green function to the left and to the right of the
contact, we use the transfer matrix of the contact $
\tilde{g}_{nsms^{\prime}}^{\sigma \sigma ^{\prime }}(x=x_{2},x^{\prime
})= \sum_{ls^{\prime \prime},\sigma^{\prime \prime}} M_{nsls^{\prime
\prime}}^{\sigma \sigma^{\prime \prime}}\tilde{g}_{ls^{\prime
\prime}ms^{\prime}}^{\sigma^{\prime \prime} \sigma ^{\prime
}}(x=x_{1},x^{\prime })$ and similarly for the $x^{\prime
}$-coordinate. Hence
\begin{equation}
\tilde{g}_{2}=M\tilde{g}_{1}M^{\dagger } \, ,
\label{leftright} 
\end{equation}
where $\tilde{g}_{2(1)}=\tilde{g}(x=x_{2(1)},x^{\prime }=x_{2(1)})$.
Up to this point the Keldysh Green functions have been obtained
exactly for the Hamiltonian (\ref{ham}). Now we should include the
proper boundary conditions to uniquely define the Green functions.  To
this end we introduce the assumptions of isotropizations at the
nodes. The incoming modes are assumed to take their quasi-classical
values described by the quasi-classical Green functions $\bar{G}$,
whereas the outgoing modes are determined by the properties of the
contact:\cite{Nazarov94:1420}
\begin{eqnarray}
\label{isotrop_1a}
\left( \bar{\Sigma}_{z}+\bar{G}_{1}\right) \left(
\bar{\Sigma}_{z}-\bar{g} _{1}\right) &=&0 \\
\label{isotrop_1b}
\left( \bar{\Sigma}_{z}+\tilde{g}_{1}\right) \left(
\bar{\Sigma}_{z}-\bar{G} _{1}\right) &=&0 \\
\label{isotrop_2a}
\left( \bar{\Sigma}_{z}-\bar{G}_{2}\right) \left(
\bar{\Sigma}_{z}+\tilde{g} _{2}\right) &=&0 \\
\label{isotrop_2b}
\left( \bar{\Sigma}_{z}-\tilde{g}_{2}\right) \left(
\bar{\Sigma}_{z}+\bar{G} _{2}\right) &=&0 \, ,
\end{eqnarray}
where $\bar{G}_{1}$ ($\bar{G}_{2}$) is the isotropic Green function in
reservoir $1$ ($2$):
$$
\left( \bar{G}_{1}\right) _{nsms'}^{\alpha \beta}=\delta
_{n,m}\delta^{\alpha \beta} (\check{G}_{1} )_{ss^{\prime}} \, .
$$
In a normal metal, the homogeneous retarded quasi-classical Green
function is
$$
\bar{G}_{R}=\left( 
\begin{array}{cc}
G_{R}^{++} & G_{R}^{+-} \\ G_{R}^{-+} & G_{R}^{--}
\end{array}
\right) =\left( 
\begin{array}{cc}
1 & 0 \\ 
0 & 1
\end{array}
\right)  \, .
$$
and the advanced quasi-classical Green function is
$\bar{G}_{A}=-\bar{G}_{R}$.  (Note that here $1$ means a unit matrix
in the basis of the transverse modes and spin, $1 \rightarrow
\delta_{ns,ms'} \hat{1}$). The Keldysh component of the Green function
is
$$
\bar{G}_{K,1(2)}=\hat{h}_{1(2)}\left(
\begin{array}{cc}
1 & 0 \\ 
0 & 1
\end{array}
\right)  \, ,
$$
where the $2 \times 2$ distribution matrix $\hat{h}$ is related to the
(non-equilibrium) distribution functions $\hat{f}(\epsilon)_{1(2)}$ in
the nodes by
$$
\hat{h}_{1(2)}=2(2\hat{f}(\epsilon)_{1(2)}-1) \, .
\label{hf}
$$

The isotropization conditions (\ref{isotrop_1a}), (\ref{isotrop_1b}),
(\ref{isotrop_2a}) and (\ref{isotrop_2b}) relate the retarded and
advanced Green function on the left and right hand side of the
contact:
\begin{eqnarray*}
\label{ret}
\tilde{g}_{R,1} &=&\left(
\begin{array}{cc}
1 & 0 \\ 
\tilde{g}_{R,1}^{-+} & 1
\end{array}
\right) \\
\tilde{g}_{R,2} &=&\left(
\begin{array}{cc}
1 & \tilde{g}_{R,2}^{+-} \\
0 & 1
\end{array}
\right) 
\end{eqnarray*}
where $\tilde{g}_{R,1}^{-+}$ and $\tilde{g}_{R,2}^{+-}$ are determined
by the Green function on the right and the scattering properties of
the contact.  The advanced Green function is related to the retarded
Green function by
$$
\tilde{g}_{A}=-\tilde{g}_{R}^{\dag} \, .
$$
The Keldysh component on the left side is determined by
(\ref{isotrop_1a}), (\ref{isotrop_1b}), (\ref{isotrop_2a}),
(\ref{isotrop_2b}) and (\ref{ret}) and is dictated by the boundary
conditions given by the isotropization process to be
\begin{eqnarray}
\tilde{g}_{K,1} &=&\left(
\begin{array}{cc}
\hat{h}_{1} 1 & \hat{h}_{1}r^{\dagger } \\
r \hat{h}_{1} & \tilde{g}_{K,1}^{--}
\end{array}
\right) \, , \\
\tilde{g}_{K,2} &=&\left(
\begin{array}{cc}
\tilde{g}_{K,2}^{++} & r^{\prime } \hat{h}_{2} \\ \hat{h}_{2}
r^{\prime \dagger } & \hat{h}_{2} 1
\end{array}
\right) \, .
\label{Kcomp}
\end{eqnarray}
We now use the relation between the Green function on the left hand
side and the right hand side of the contact (\ref{leftright}) to
obtain the retarded Green functions
\begin{eqnarray*}
\tilde{g}_{R,1}^{-+} & = & 2r \\
\tilde{g}_{R,2}^{+-} & = & 2r' \, ,
\end{eqnarray*}
and the Keldysh Green function
\begin{eqnarray}
\tilde{g}_{K,1}^{--} & = & t' \hat{h}_2 t'^{\dag} + \hat{h}_1 r^{\dag} \\
\tilde{g}_{K,2}^{++} & = & t \hat{h}_1 t^{\dag} + r' \hat{h}_2 r'^{\dag} \,
.
\label{Keld}
\end{eqnarray}
Inserting the expression the Keldysh component (\ref{Keld}) into
(\ref{current}) and using (\ref{hf} )we finally find the expression
for the current through the contact (\ref{curLB}).

\section{Contact conductances}
\label{s:contact}

The four conductance parameters $G^{\uparrow}$, $G^{\downarrow}$,
$\mbox{Re} G^{\uparrow \downarrow}$ and $\mbox{Im} G^{\uparrow
\downarrow}$ depend on the microscopic details as illustrated below
for 3 elementary model contacts: a diffusive, a ballistic, and a
tunnel contact.

\subsection{Diffusive contact}

We consider first a diffusive contact between a normal metal node and
a ferromagnetic node and show the relations between the conductances
(see Fig.~\ref{f:diffusive}). The cross-section of the contact is $A$,
the length of the normal metal part of the contact is $L^{N}$, the
length of the ferromagnetic part of the contact is $L^{F}$, the
conductivity on the normal side $\sigma ^{N}$ and the spin-dependent
conductivities on the ferromagnetic side $\sigma ^{Fs}$. The
conductance of the normal part is $G^{N}_D=A\sigma ^{N}/L^{N}$ and the
spin-dependent conductances of the ferromagnetic part are
$G^{Fs}_D=A\sigma ^{Fs}/L^{F}$. The spin-dependent conductances of the
whole contact are obtained simply as the diffusive ferromagnetic and
normal metal regions in series:
\begin{eqnarray}
G^{\uparrow}_D & = &
\frac{G^{F\uparrow}_DG^{N}_D}{G^{F\uparrow}_D+G^{N}_D} \, , \\
G^{\downarrow}_D & =
&\frac{G^{F\downarrow}_DG^{N}_D}{G^{F\downarrow}_D+G^{N}_D}.
\end{eqnarray}
These spin-dependent conductances ($G^{\uparrow}_D$ and
$G^{\downarrow}_D$) fully describe collinear transport (in the absence
of spin-flip scattering). For non-collinear magnetizations the mixing
conductance Is also needed. It can be derived from the scattering
matrix, {\em e.g.} with the method developed in
Ref.~\cite{Nazarov94:134}. Here we use a much simpler approach
based on the diffusion equation, describing the scattering properties
of the contact by a spatially dependent distribution matrix. The
current density on the normal side of the contact ($x<0$) is
$\hat{\imath}(x<0)=\sigma ^{N}\partial _{x}\hat{f}$ and consequently
the total current is
$$
\hat{I}(x<0)=G^{N}_D (L^N \partial _{x}) \hat{f} \, ,
$$
where $\hat{f}$ is the spatially dependent distribution matrix on the normal
side in the contact. In the normal metal node the boundary condition is
\begin{equation}
\hat{f}(x=-L^{N})=\hat{f}^{N}.  \label{normbound}
\end{equation}
In a ferromagnet spin-up and spin-down states are incoherent, and
hence spins non-collinear to the magnetization direction relax
according to (\ref{spinprecession}) and only spins collinear with the
magnetization will propagate sufficiently far away from the
NF-interface. We assume that the ferromagnet is sufficiently strong
and that the contact is longer than the ferromagnetic decoherence
length $\xi=\sqrt{D/h_{\mbox{ex}}}$, where $D$ is the diffusion
constant and $h_{\mbox{ex}}$ is the exchange splitting.  The
decoherence length is typically very short in ferromagnets, $\xi$=2nm
in Ni wires\cite{Petrashov99:3281}. The distribution function on the
ferromagnetic side can then be represented by a 2-component
distribution function
$$
\hat{f}(x>0)= \hat{u}^{\uparrow} f^{\uparrow }+ \hat{u}^{\downarrow}
f^{\downarrow } \, ,
$$
where $\hat{u}^{\uparrow}$ and $\hat{u}^{\downarrow}$ are the
spin-projection matrices (\ref{spinproj_up}) and (\ref{spinproj_do}).

We allow for a spin-accumulation collinear to the magnetization
direction in the ferromagnet. The boundary condition determined by the
distribution function in the ferromagnetic node is thus
\begin{eqnarray}
\label{ferbound_up}
f^{\uparrow}(x =L^{F}) &=& f^{F\uparrow } \, , \\
\label{ferbound_do}
f^{\downarrow }(x=L^{F}) & = &f^{F\downarrow } \, .
\end{eqnarray}
The total current in the ferromagnet is
$$
\hat{I}(x>0)= G^{F\uparrow}_D \hat{u}^{\uparrow} \partial
_{x}f^{\uparrow }+ G^{F\downarrow}_D \hat{u}^{\downarrow} \partial
_{x}f^{\downarrow } \, .
$$
We assume that the resistance of the diffusive region of the contacts
is much larger than the contact resistance between the normal and the
ferromagnetic metal. The distribution function is in this limit
continuous across the normal metal - ferromagnetic interface,
\begin{equation}
\hat{f}(0^{+})=\hat{f}(0^{-})  \label{intbound} \, .
\end{equation}
Current conservation on the left ($x < 0$) and on the right ($x > 0$)
of the normal metal-ferromagnet interface dictates
\begin{equation}
\partial_{x}\hat{I}=0  \label{curconsdiff}  \, .
\end{equation}
Note that the component of the spin-current that is non-collinear to
the magnetization direction in the ferromagnet is not conserved on
going through the interface due to strong relaxation induced by
(\ref{spinprecession}).  The first order differential equation
(\ref{curconsdiff}) and the boundary conditions (\ref{normbound}),
(\ref{ferbound_up}), (\ref{ferbound_do}) and (\ref{intbound}) uniquely
determine the distribution functions and hence the conductance in the
diffusive contact. The current on the normal side of the contact
becomes
\begin{eqnarray}
e \hat{I} & = & G^{\uparrow}_D \hat{u}^{\uparrow}
(\hat{f}^F-\hat{f}^N) \hat{u}^{\downarrow} + G^{\downarrow}_D
\hat{u}^{\downarrow} (\hat{f}^F-\hat{f}^N) \hat{u}^{\downarrow}
\nonumber \\
& + & G^{N}_D \left[ \hat{u}^{\uparrow} (\hat{f}^F-\hat{f}^N)
\hat{u}^{\downarrow} + \hat{u}^{\uparrow} (\hat{f}^F-\hat{f}^N)
\hat{u}^{\downarrow} \right] \, .
\label{curdiff}
\end{eqnarray}
The current in a diffusive contact thus takes the generic form
(\ref{curgen}) with $G^{\uparrow}=G^{\uparrow}_D$,
$G^{\downarrow}=G^{\downarrow}_D$ and $G^{\uparrow
\downarrow}=G^{N}_D$. The mixing conductance is thus real and only
depends on the normal conductance. The latter results can be
understood as a consequence of the effective spin-relaxation of spins
non-collinear to the local magnetization direction. Those spins cannot
propagate in the ferromagnet, and consequently the effective
conductance can only depend on the conductance in the normal metal as
(\ref{curdiff}) explicitly demonstrates.

\subsection{Ballistic contact}

A simplified expression for the conductances can be found for a
ballistic contact. Firstly, the reflection and transmission
coefficients appearing in (\ref{Gup}), (\ref{Gdo}) and (\ref{Gmixing})
are diagonal in the space of the transverse channels since the
transverse momentum is conserved. In a simplified
model\cite{Bauer92:1676} the transmission channels are either closed
$t=0$ or open $t=1$. The conductances (\ref{Gup}), (\ref{Gdo}) and
(\ref{Gmixing}) can then be found by simply counting the number of
propagating modes.  We obtain the spin-dependent conductances
\begin{eqnarray}
G^{\uparrow}_B & = & \frac{e^2}{h} N^{\uparrow} \, \\
G^{\downarrow}_B & = & \frac{e^2}{h} N^{\downarrow} \, ,
\end{eqnarray}
where $N^{\uparrow}$ is the number of spin-up propagating channels and
$N^{\downarrow}$ is the number of spin-down propagating channels.  The
mixing conductance is determined by
$$
G^{\uparrow\downarrow}_B=\mbox{max} (G^{\uparrow}_B,G^{\downarrow}_B)
$$
and is real. In a quantum mechanical calculation the channels just
above the potential step are only partially transmitting and the
channels below a potential step can have a finite transmission
probability due to tunneling. Furthermore, the band structure of
ferromagnetic metals is usually complicated and interband scattering
exists even at ideal interfaces. We may therefore expect that in
general the phase of the scattered wave will be relevant giving a
non-vanishing imaginary part of the mixing
conductance. First-principles calculations of the complete conductance
matrix are therefore highly desirable.

\subsection{Tunnel contact}

For a tunneling contact the transmission coefficients are
exponentially small and the reflection coefficients have a magnitude
close to one. The spin-dependent conductances are
\begin{eqnarray}
G^{s}_T & = & \frac{e^2}{h} \sum_{nm} |t^{nm}_s|^2 \, .
\label{spinGtun}
\end{eqnarray}
For simple models of tunnel barriers $r^{nm}_s=\delta^{nm}
\exp{i\phi^n} - \delta r^{nm}_s$, where the phase-shift $\phi^n$ is
spin-independent. We expand (\ref{Gmixing}) in terms of the small
correction $\delta r^{nm}_s$ and find that
$$
\mbox{Re} G^{\uparrow \downarrow}_T =
(G^{\uparrow}_T+G^{\downarrow}_T)/2 \, ,
$$
where $G^{\uparrow}_T$ and $G^{\downarrow}_T$ are the spin-dependent
tunneling conductances (\ref{spinGtun}). Since the transmission
coeeficients in a tunnel contact are all exponentially small, the
imaginary part of $G^{T\uparrow \downarrow}$ is of the same order of
magnitude as $G^{T\uparrow}$ and $G^{T\downarrow}$ but it is not
universal and depends on the details of the contact.

\section{Illustrations of the theory}
\label{s:illustrations}

We will in this chapter illustrate the appeal of the circuit theory of
spin-transport by computing the transport properties of two-terminal
devices and a three terminal device. It is assumed that the normal
metal node in these devices is smaller than the spin-diffusion length
so that the spatial distribution function is homogeneous within the
node. 

\subsection{Two terminals}
\label{s:two}

First we consider a normal metal node attached to two ferromagnetic
reservoirs with identical contacts , {\em e.g.}
$G^{\uparrow}_1=G^{\uparrow}_2=G^{\uparrow}$,
$G^{\downarrow}_1=G^{\downarrow}_2=G^{\downarrow}$ and $G^{\uparrow
\downarrow}_1=G^{\uparrow \downarrow}_2=G^{\uparrow \downarrow}$.  The
relative angle between the magnetization in the two ferromagnetic
reservoirs is $\theta$. With the aid of (\ref{curcons}) and
(\ref{curgen}) we find the current
\begin{equation}
I(\theta) = \frac{G}{2} V \left(1- \frac{P^2}{1+g_{\mbox{sf}}}
\frac{\tan^2{\theta/2}}{\tan^2{\theta/2}+\alpha} \right)
\, ,
\label{twocur}
\end{equation}
where 
\begin{equation}
\alpha = \frac{|\eta|^2+g_{\mbox{sf}}(3\eta_R+2
g_{\mbox{sf})}}{(\eta_R+2g_{\mbox{sf}})(1+g_{\mbox{sf}})} \, .
\end{equation}
Here, we have introduced the total conductance of one contact
$G=G^{\uparrow}+G^{\downarrow}$, the polarization
$P=(G^{\uparrow}-G^{\downarrow})/(G^{\uparrow}+G^{\downarrow})$, the
relative mixing conductance $\eta=2G^{\uparrow
\downarrow}/(G^{\uparrow}+G^{\downarrow})$ and the ratio of the
'spin-flip conductance'\cite{Brataas99:93} $G_{\mbox{sf}}=e^2/(2\delta
\tau_{\mbox{sf}})$ ($\delta$ is the energy level spacing) to the
conductance of the whole device in the parallel configuration
$g_{\mbox{sf}}=G_{\mbox{sf}}/(G/2))$. The current is an even function
of $\theta$. Note that $\alpha>1$. When the magnetizations are
parallel ($\theta=0$), there is no spin-accumulation on the normal
metal node and the current is given by Ohm's law
$I_{\mbox{P}}=I(\theta=0)=GV/2$. The anti-parallel magnetization
configuration ($\theta=\pi$) generates the largest spin-accumulation,
reducing the particle current to
$I_{\mbox{AP}}=I(\theta=\pi)=G[1-P^2/(1+g_{\mbox{sf}})]V/2$. In this
case the magneto-resistance ratio
$(I_{\mbox{P}}-I_{\mbox{AP}})/I_{\mbox{P}}$ is
$P^2/(1+g_{\mbox{sf}})$, irrespective of the relative mixing
conductance $\eta$. Naturally, the spin-accumulation and consequently
the magnetoresistance decreases with spin-flip relaxation time. Any
spin flip reduces the effective polarization. The result for long
spin-flip relaxation times ($g_{\mbox{sf}} \ll 1$) was previously
obtained for two tunnel junctions\cite{Brataas99:93} and we have thus
generalized it to arbitrary contacts. For general $\theta$ the current
depends on $\alpha$ and thus on the mixing conductance. In
Fig.~\ref{f:two} we plot the current vs. the relative magnetization
angle $\theta$ for a given effective polarization
$P^2/(1+g_{\mbox{sf}})$ and a number of values of $\alpha$
($\alpha=1$, $\alpha=10$ and $\alpha=100$).  In general, the current
increases with increasing $\eta$. As one can see from (\ref{curgen}) a
large relative mixing conductance means that spins orthogonal to the
magnetization in the reservoirs easily can escape from the normal
metal node. This suppresses the spin accumulation. Therefore when
$\eta \gg 1$ and $\theta$ is not close to $\pi$ the current approaches
Ohm's law $I=GV/2$. Except for the anti-parallel magnetizations, the
angle dependence vanishes in that limit which explains the sharp dip
at $\theta$ close to $\pi$.

Fig.~\ref{f:two} illustrates a universal property of the two-terminal
device with non-collinear magnetizations that is independent of the
contact conductances and the spin-flip relaxation time. By scaling the
total conductance modulation to the difference between the conductance in
the parallel and anti-parallel configuration
$(G/2)P^2/(1+g_{\mbox{sf}})$ the current change for {\em any} two
terminal device with a diffusive normal metal node should be above the
universal curve determined by the minimum value of $\eta$,
$|\eta|=1$. Thus, according to our theory the current
vs. magnetization angle relation for a spin valve must lie above the
universal curve obtained for $|\eta|=1$.

The result (\ref{twocur}) has been derived for normal metals islands
that can be described as Fermi liquids\cite{Brataas9906065}. It was
recently generalized to transport through a Luttinger liquid by
Balents and Egger\cite{Balents00:3463}. In a Luttinger liquid the
current is non-linear in the applied source-drain bias voltage, and
the spin-charge separation reduces the
spin-accumulation\cite{Balents00:3463}. Coulomb charging effects can
also reduce the spin-accumulation in the linear response
regime\cite{Brataas99:93}.

\subsection{Three terminals}
\label{s:three}

Let us now consider a set-up similar to the Johnson spin-transistor as
shown in Fig.~\ref{f:threeJohnson} which was also discussed by Geux
{\em et al.}~\cite{Geux00:119} within the context of a multi-terminal
Landauer-B\"{u}ttiker formalism for collinear magnetization
configurations only.

A small normal metal node is attached to two ferromagnetic reservoirs
and one normal metal reservoir by three contacts. A voltage bias
applied to the ferromagnetic reservoir F1 and the normal metal
reservoir N causes a current between the same reservoirs passing
through the normal metal node. The spin-accumulation on the normal
metal node injected by F1 affects the chemical potential of
ferromagnet F2, which is adjusted such that the charge current into F2
vanishes. We characterize the contact between the first (second)
ferromagnet and the normal metal node by the total conductance
$G_1=G_{1}^{\uparrow}+G_{1}^{\downarrow}$
($G_2=G_2^{\uparrow}+G_2^{\downarrow}$), the polarization
$P_1=(G_{1}^{\uparrow}-G_{1}^{\downarrow})/G_1$
($P_2=(G_{2}^{\uparrow}-G_{2}^{\downarrow})/G_2$) and the relative
mixing conductance $\eta_2=2G_2^{\uparrow
\downarrow}/(G_2^{\uparrow}+G_2^{\downarrow})$ ($\eta_1=2G_1^{\uparrow
\downarrow}/(G_1^{\uparrow}+G_1^{\downarrow})$). The contact between
the normal metal reservoir and the normal metal node is characterized
by a single conductance parameter, $G_{3N}$. $\theta$ is the relative
angle between the magnetization of ferromagnet F1 and ferromagnet
F2. We assume that the typical rate of spin-injection into the node
is faster than the spin-flip relaxation rate, so that the right hand
side of (\ref{curcons}) can be set to zero.

The current through the normal metal node is invariant with respect to
a flip in the magnetization direction of ferromagnet F1 or F2:
$I(\theta) = I(\theta+\pi)$. However the chemical potential of
ferromagnet F2 changes during the same process since it is sensitive
to the magnitude and direction of the spin-accumulation on the normal
metal node: $\mu_2(\theta) \neq \mu_2(\theta+\pi)$. A
spin-`resistance' $R_s(\theta)$ can be defined as the ratio between
the difference in the chemical potential of ferromagnetic F2 when
ferromagnet F1 or ferromagnet F2 is flipped:
$R_S=(V_2(\theta+\pi)-V_2(\theta))/I(\theta)$. In the collinear
configuration ($\theta=0$) the spin-resistance $R_s(\theta=0)$ is
thus the ratio between the difference in the chemical potential of
ferromagnetic F2 when its magnetization is parallel
($\mu_2^{\mbox{P}}$) and anti-parallel ($\mu_2^{\mbox{AP}}$) to the
magnetization of ferromagnet F1 and a current ($I$) passes from
ferromagnet F1 to the normal metal reservoir N. With the aid of the
general conductances, valid for arbitrary contacts, we solve for the
non-equilibrium distribution function on the normal metal node
(\ref{curcons}) under the condition that no particle current enters
ferromagnet F2. Using the solution for the non-equilibrium
distribution function we find the current (\ref{curgen}) through the
system and subsequently the non-equilibrium chemical potential of
ferromagnet F2.

Let us first discuss the results in the collinear configuration,
$\theta=0$ and $\pi$. The spin-resistance can be simply expressed as
\begin{equation}
R_S(\theta=0)= \frac{2P_1 P_2}{G_1
(1-P_1^2)+G_{3N} + G_2 (1-P_2^2)} \, ,
\label{spinresistance}
\end{equation}
and is independent of the relative mixing conductances $\eta_1$ and
$\eta_2$ that are only relevant for the transport properties in
systems with non-collinear magnetization configurations. The
spin-resistance is proportional to the product of the polarizations of
the contacts to ferromagnet F1 and ferromagnet F2.  In order to
measure a large effect of the spin-accumulation, {\em e.g.} a large
spin-resistance, highly resistive contacts should be used. On the
other hand, the resistance has to be small enough so that the
transport dwell time is shorter than the spin-flip relaxation. The
simple result (\ref{spinresistance}) covers a large class of
experiments, since we have not specified any details about the
contacts between the reservoirs and the normal metal node. It is
noted, though, that Eq.~(\ref{spinresistance}) is only valid for a
normal metal node that is smaller than the spin-diffusion length and
can therefore not be applied directly to Johnson's
experiment\cite{Johnson85:1790}.

We can understand that the present results are quite different from
those of Ref.~\cite{Geux00:119} as follows. The general
formulation in terms of transmission probabilities of Geux {\em et
al.} is exact. However, in order to include the effects of
spin-relaxation the transmission probabilities were treated as
pair-wise resistors between the reservoirs. This corresponds to an
equivalent circuit in which resistors connect the three reservoirs in
a ``ring" topology.  The present model, on the other hand, can be
described by a ``star" configuration circuit, in which all resistors
point from the reservoirs to a single node.  The present model is more
accurate when the contacts dominate the transport properties, whereas
Geux's model is preferable when the resistance of the normal metal
island is important. Effectively, Johnson's thin film device appears
to be closer to the star configuration.

Let us now proceed to discuss the results when the magnetization
directions are non-collinear. The analytical expression for the
spin-resistance is much simpler when the two contacts F1-N and F2-N
are identical, $G_1=G_2\equiv G$, $P_1=P_2 \equiv P$ and
$\eta_1=\eta_2 \equiv \eta$. Furthermore we disregard the imaginary
part of the mixing conductance which is very small or zero in the
model calculations of tunnel, ballistic and diffusive contacts
presented in this paper as well as in recent first-principle
band-structure calculations\cite{Xia01}. The spin-resistance then has
the simple form
\begin{equation}
R_S = \frac{2(G_{3N}+ 2 G \eta) P^2
\cos(\theta)}{(G_{3N}+G(1+\eta-P^2))^2-G^2\cos^2(\theta)(1-\eta
-P^2)^2}
\end{equation}
The spin-resistance is an even function of the relative angle
between the magnetization directions $\theta$ and we recover the
result (\ref{spinresistance}) when $\theta=0$. The spin-resistance
vanishes when the magnetizations are perpendicular $\theta=\pi/2$ as
expected from the symmetry of the systems. The angular dependence is
approximately proportional to $\cos(\theta)$ when the relative mixing
conductance is not too large, $\eta \approx 1$. For larger mixing
conductances $\eta \gg 1$ the spin-accumulation on the normal metal
island is strongly suppressed in the perpendicular configuration
$\theta=\pi/2$ due to the large transport rates for spins between the
normal metal node and ferromagnet F2. Consequently the
spin-resistance is small and only weakly dependent on the relative
angle around $\theta=\pi/2$.

Another novel three-terminal device comprising of three ferromagnetic
reservoirs (the ``spin-flip transistor"), which utilizes the added
functionality provided by non-collinear magnetization directions, was
introduced in Ref.~\cite{Brataas00:2481}.

\section{Conclusion}
\label{s:con}

We developed a mesoscopic circuit theory of spin-transport in
multi-terminal hybrid ferromagnet - normal metal systems starting from
microscopic principles. Based on conservation of spin-current on each
node the circuit theory is parameterized by the conductances of the
contacts, {\em viz.} two spin-dependent conductances and a (complex)
mixing conductance. The latter is a novel concept relevant for
transport in systems with non-collinear magnetization
configurations. Explicit expressions for the conductances for
diffusive, ballistic and tunnel contacts have been derived. The
circuit theory leads to simple and quite general results for the
conductance of two-terminal and three-terminal devices, like Johnson's
spin-transistor. For two-terminal systems a universal lower limit for
the current modulation as a function of the relative magnetization has
been found.

After completion of this work a complementary approach to
spin-transport in ferromagnetic-normal metal systems starting from the
scattering matrices was presented in Ref.\
\cite{Waintal00:12317}. Using random matrices to describe the
scattering within disordered normal metal nodes equivalent results to
our (\ref{curLB}) were obtained.

\begin{acknowledgement}
We would like to thank W.\ Belzig, P.\ W.\ Brouwer, B.\ I.\ Halperin,
D.\ Huertas Hernando, J.\ Inoue and X.\ Waintal for discussions.

A.\ B.\ is financially supported by the Norwegian Research
Council. This work is part of the research program for the ``Stichting
voor Fundamenteel Onderzoek der Materie'' (FOM), which is financially
supported by the ``Nederlandse Organisatie voor Wetenschappelijk
Onderzoek'' (NWO). We acknowledge benefits from the TMR Research
Network on ``Interface Magnetism'' under contract No. FMRX-CT96-0089
(DG12-MIHT) and support from the NEDO joint research program
(NTDP-98).
\end{acknowledgement}

\appendix

\section{Transformation from a non-collinear to a collinear configuration}
\label{s:transform}

We consider the transmission and reflection matrices between a normal
metal and a ferromagnet. The Schr\"{o}dinger equation is
$$
\hat{H}({\bf r}) \psi ({\bf r})=E\psi ({\bf r}),
$$
where $\psi ({\bf r})$ is a two-component spinor. The Hamiltonian is
\begin{equation}
\hat{H}({\bf r})=\hat{U}\left[ -\frac{1}{2m}\nabla
^{2}\hat{1}+V_{s}({\bf r}) \sigma_z +\hat{V}_{c}({\bf r})\right]
\hat{U} \, .
\label{HamNF}
\end{equation}
We consider transport between a normal metal and a uniform
ferromagnet, so that the magnetization direction ${\bf m}$ is a
spatially independent unit vector. In (\ref{HamNF}) $V_{s}({\bf r})$
denotes the spin-dependent potential and $\hat{V}_{c}({\bf r})$ is the
scattering potential of the contact.  The direction of the
magnetization is represented by the angles $\theta $ and $\varphi $ as
${\bf m=(\sin \theta \cos \varphi ,\sin \theta \sin \varphi ,\cos
\theta )}$. The hermitian and unitary matrix $\hat{U}$ that
diagonalizes the spin-dependent potential is
$$
\hat{U}=\left( 
\begin{array}{cc}
\cos( \theta /2) & \sin(\theta /2)e^{-i\varphi/2} \\
 \sin(\theta /2) e^{i\varphi/2} & -\cos(\theta /2)
\end{array}
\right) .
$$
The spin-dependent potential vanishes in the normal metal, $V_{s}({\bf
r})=0$ for $x<x_{l}$ and attains a constant value in the ferromagnet
$V_{s}({\bf r} )=V_{s}$ for $x>x_{r}$. The contact is represented by
the scattering potential
\begin{equation}
\hat{V}_{c}({\bf r})=\left(
\begin{array}{cc}
V_{\uparrow }({\bf r}) & V_{\mbox{sf}}({\bf r}) \\
V_{\mbox{sf}}^{\dagger }({\bf r}) & V_{\downarrow }({\bf r})
\end{array}
\right) 
\label{Vc}
\end{equation}
and attains the bulk values within the normal metal and the
ferromagnet for $x<x_l$ and $x>x_r$,respectively. The off-diagonal
terms in (\ref{Vc}) represent the exchange potentials due to a
non-collinear magnetization in the contact, spin-orbit interaction or
spin-flip scatterers. The Hamiltonian (\ref{HamNF}) can be
diagonalized in spin-space by
\begin{equation}
\psi ({\bf r})=\hat{U} \phi ({\bf r}) \, .  \label{gauge}
\end{equation}
The Schr\"{o}dinger equation for the spinor $\phi ({\bf r})$ is
$$
\left[ -\frac{1}{2m}\nabla ^{2}\hat{1}+V_{s}({\bf r})\sigma
_{z}+\hat{V}_{c}( {\bf r})-E\right] \phi ({\bf r})=0 \, .
$$
Let us now consider an incoming wave from the normal metal in the
transverse mode $n$ and with spin $s$ collinear to the magnetization
in the ferromagnet. The wave function in the normal metal is
\begin{eqnarray}
\phi _{s}^{n}({\bf r})=&& \sum_{ms^{\prime }}\frac{\chi_{N}^{m}({\bf
\rho })}{ \sqrt{k^{m}}} \times \nonumber \\
 && [ \delta _{s^{\prime }s}\delta ^{mn}\xi_{s}e^{ik^{n}x}+
r_{\mbox{c},s^{\prime }s}^{mn}\xi_{s^{\prime }} e^{-ik^{m}x} ] ,
\end{eqnarray}
where $\xi_{\uparrow }^{\dagger }=\left( 1,0\right) $ and
$\xi_{\downarrow }^{\dagger }=\left( 0,1\right) $ are the spinors,
$\chi_{N}^{m}( {\bf \rho })$ is the transverse wave function, $k^{m}$
is the longitudinal wave vector for mode $m$ and
$r_{\mbox{c},s^{\prime }s}^{mn}$ is the reflection matrix from state
$ns$ to state $ms^{\prime }$. We would like to transform the result
for the reflection matrix into a basis with arbitrary spin
quantization axis. To this end we introduce the incoming spinor wave
function
\begin{eqnarray}
\psi_{s}^{n}({\bf r})= && \sum_{ms^{\prime }}\frac{\chi_{N}^{m}({\bf
\rho })}{ \sqrt{k^{m}}} \times \nonumber \\
&& \left[ \delta _{s^{\prime }s}\delta ^{mn}\xi_{s}
e^{ik^{n}x}+r_{s^{\prime }s}^{mn}\xi_{s^{\prime }} e^{
-ik^{m}x}\right] \, .
\label{psiloc}
\end{eqnarray}
Using the transformation (\ref{gauge}) we can also write the wave
function spinor in terms of the basis states $\phi ({\bf r})$ as
\begin{equation}
\psi_{s}^{n}({\bf r})=\hat{U} \sum_{\sigma }\phi_{\sigma
}^{n}({\bf r} )a_{\sigma s}, \label{psiexpan}
\end{equation}
where $a_{\sigma s}$ are expansion coefficients to be determined by equating
(\ref{psiloc}) and (\ref{psiexpan}). We thus find that
$$
a_{\sigma s}=\xi_{\sigma }^{\dagger }U\xi_{s}=U_{\sigma s}
$$
and 
\begin{equation}
r_{s^{\prime }s}^{mn}=\sum_{\sigma ^{\prime }\sigma }U_{s^{\prime
}\sigma ^{\prime }} r_{\mbox{c},\sigma ^{\prime }\sigma
}^{mn}U_{\sigma s}.
\label{reftrans}
\end{equation}
Disregarding spin-flip processes in the contact the transformation of
the reflection matrix can be written as
$$
\hat{r}^{nm}=\hat{u}_{\uparrow }r_{\mbox{c,}\uparrow \uparrow
}^{mn}+\hat{u} _{\downarrow }r_{\mbox{c,}\downarrow \downarrow }^{mn},
$$
where the spin-projection matrices $\hat{u}_{\uparrow }$ and
$\hat{u}_{\downarrow }$ are defined in (\ref{spinproj_up}) and
(\ref{spinproj_do}). Spin-flip processes can also be included by using
the general transformation (\ref{reftrans}), but the reflection matrix
$\hat{r}^{nm}$ can then not be expressed in terms of the
spin-projection matrices only.

We can perform a similar calculation in order to find the
transformation of the transmission coefficients from the ferromagnet
into the normal metal. In the basis where the spin-quantization axis
is collinear with the magnetization the incoming wave from the
ferromagnet is
\begin{eqnarray}
\phi _{s}^{n}({\bf r})= && \sum_{ms^{\prime }}\frac{\chi_{Fs}^{m}({\bf
\rho })}{ \sqrt{k_{s}^{m}}} \times \nonumber \\
&&\left[ \delta _{s^{\prime }s}\delta ^{mn}\xi_{s}
e^{ik_{s}^{n}x}+r_{\mbox{cF},s^{\prime }s}^{mn}\xi_{s^{\prime }}
e^{-ik_{s}^{m}x} \right] ,
\end{eqnarray}
where $\chi_{m}^{Fs}({\bf \rho })$ is the (spin-dependent) transverse
wave function and $k_{s}^{m}$ is the spin-dependent Fermi
wave-vector. The outgoing wave into the normal metal is
$$
\phi _{s}^{n}({\bf r})=\sum_{ms^{\prime }}\frac{\chi_{N}^{m}({\bf \rho
})}{ \sqrt{k^{m}}}t_{\mbox{cF},s^{\prime }s}^{m,n}\xi_{s^{\prime
}}\exp (ik_{s}^{m}x).
$$
By transforming the outgoing wave into an arbitrary magnetization
direction according to (\ref{gauge}), we see that the transmission
coefficient from a state with spin $s$ collinear to the magnetization
direction in the ferromagnet to a state with spin $s^{\prime }$
collinear to the spin-quantization axis along the $z$-direction is
$$
\hat{t}^{\prime nm}=\hat{U}\hat{t}_{\mbox{cF}}^{nm}.
$$
In the absence of spin-flip scattering in the contact, we have
$$
t_{ss^{\prime }}^{\prime nm}=U_{ss^{\prime }}t_{\mbox{cF},s^{\prime
}}^{nm}.
$$
The current in the normal metal is (for a given energy shell) (note
that we associate the first index with $\Psi $ and the second index
with $\Psi ^{\dagger }$)
\begin{eqnarray}
\frac{h}{e} \hat{I} & = & M\hat{f}^{N} \nonumber \\
&& -\sum_{nm}\left[ \hat{r}^{mn}\hat{f}^{N}\left( \hat{r}^{nm}\right)
^{\dagger }-\hat{t}^{\prime mn}\hat{f}^{F}\left( \hat{t}^{\prime
nm}\right) ^{\dagger }\right] \, .
\end{eqnarray}
The contribution from the transmission probability to the spin-current
is therefore
\begin{eqnarray*}
eI_{\alpha \delta }^{F} & = & \frac{e^2}{h} \sum_{nm\beta }t_{\alpha
\beta }^{\prime mn}f_{\beta }^{F}\left( t_{\beta \delta }^{\prime
nm}\right) ^{\dagger } \nonumber \\
& = & \frac{e^2}{h} \sum_{nm\beta }U_{\alpha \beta }f_{\beta
}^{F}\left| t_{\beta }^{\prime mn}\right| ^{2}U_{\beta \delta }
\end{eqnarray*}
Whereas the contribution from the transmission probability becomes
$$
e \hat{I}^{F}=G_{\uparrow }\hat{u}_{\uparrow }f_{\uparrow
}^{F}+G_{\downarrow }\hat{u}_{\downarrow }f_{\downarrow }^{F} \, ,
$$
where the spin-dependent conductance is
$$
G_{s}=\frac{e^{2}}{h}\sum_{nm}\left| t_{s}^{\prime mn}\right| ^{2}.
$$
Similarly, the contribution from the normal metal is
\begin{eqnarray*}
e\hat{I} &=&M \hat{f}^{N}-\sum_{ss^{\prime
}}\hat{u}^{s}\hat{f}^{N}\hat{u}^{s^{\prime }}r_{s}^{mn}\left(
r_{s^{\prime }}^{mn}\right) \\
&=&-G_{\uparrow }\hat{u}_{\uparrow }\hat{f}^{N}\hat{u}_{\uparrow
}-G_{\downarrow }\hat{u}_{\downarrow }\hat{f}^{N}\hat{u}_{\downarrow
}- \nonumber \\
&& G_{\uparrow \downarrow }\hat{u}_{\uparrow
}\hat{f}^{N}\hat{u}_{\downarrow }-G_{\uparrow \downarrow }^{\ast
}\hat{u}_{\downarrow }\hat{f}^{N}\hat{u} _{\uparrow } \, ,
\end{eqnarray*}
where we have used the unitarity of the scattering matrix so that $
M-\sum_{nm}\left| r_{s}^{nm}\right| ^{2}=\sum_{nm}\left| t_{s}^{\prime
nm}\right| ^{2}$ and the mixing conductance is introduced as
$$
G_{\uparrow \downarrow }=\frac{e^{2}}{h}\left[ M-\sum_{nm} \left(
r_{\uparrow }^{mn} \right)^{*} r_{\downarrow }^{nm}\right] \, .
$$

\begin{figure}
\caption{A many-terminal circuit consisting of normal metals (N) and
ferromagnets (F) with arbitrary magnetization directions. The normal
metals (ferromagnets) are coupled to adjacent nodes by contacts which
determine the resistance of the system.}
\label{f:circuit}
\end{figure}

\begin{figure}
\caption{A contact between a ferromagnetic node and a normal metal
node. The current is evaluated at the normal metal side (dotted
line). The transmission coefficient from the ferromagnet to the normal
metal is $t'$ and the reflection matrix from the normal metal to the
normal metal is $r$.}
\label{f:contact}
\end{figure}

\begin{figure}
\caption{A diffusive contact between a ferromagnetic node and a normal
metal node. The length of the ferromagnetic (normal metal) part of the
contact is $L^F$ ($L^N$), and the conductivity of the ferromagnetic
(normal metal) is $\sigma^F$ ($\sigma^N$). The cross-section of the
contact is $A$.}
\label{f:diffusive}
\end{figure}

\begin{figure}
\caption{The two-terminal device comprising a normal metal node
attached to two ferromagnetic reservoirs (F1 and F2) with arbitrary
relative magnetization direction $\theta$. A source-drain bias $V$ is
applied between the ferromagnetic reservoirs and a current $I$ flows
between the two reservoirs. The contact between ferromagnetic F1 (F2)
and the normal metal node is characterized by the conductances
$G_1^{\uparrow}$, $G_1^{\downarrow}$, and $G_1^{\uparrow \downarrow}$
($G_2^{\uparrow}$, $G_2^{\downarrow}$, and $G_2^{\uparrow
\downarrow}$).}
\label{f:two}
\end{figure}

\begin{figure}
\caption{The current through the two-terminal device as a function of
the relative angle $\theta$ between the magnetization directions in F1
and F2. The current is normalized by the current in the parallel
configuration $I^{\mbox{P}}=GV/2$ and reaches the minimum
$I=(1-P^2/(1+g_{\mbox{sf}}))GV/2 \equiv (1-P^2_{\mbox{eff}}) GV/2$
when the magnetizations are anti-parallel ($\theta=\pi$). The lowest
solid line is for the minimum relative mixing conductance
$|\eta=1|$. This line forms the lowest possible universal curve for
the conductance vs. relative angle $\theta$, since all other values of
the mixing conductance lies above this line.}
\label{f:tworesults}
\end{figure}

\begin{figure}
\caption{The three terminal Johnson spin-transistor. A bias voltage is
applied between a ferromagnet and a normal metal and a current flows
between the same reservoirs through a normal metal node. The potential
on another ferromagnet is measured when its magnetizations is parallel
or anti-parallel to the first ferromagnet. The contact between the
first (second) ferromagnet and the normal metal node is characterized
by the total conductance $G_1$ ($G_2$), and the polarization $P_1$
($P_2$). The conductance between the normal metal node and the normal
metal reservoir is $G_{3N}$.}
\label{f:threeJohnson}
\end{figure}


\begin{thebibliography}{99}
\bibitem{Meservey94:173} R.\ Meservey and P.\ M.\ Tedrow, Phys.\ Rep.\
\textbf{ 238}, (1994) 173.

\bibitem{Levy94:367} P.\ M.\ Levy, Sol. St. Phys. \textbf{ 47}, (1994)
367; M.\ A.\ M.\ Gijs and G.\ E.\ W.\ Bauer, Adv. Phys. \textbf{ 46},
(1997) 285.

\bibitem{Levy99:223} P.\ M.\ Levy and S.\ F.\ Zhang, Curr. Opin. Solid
St. M.  \textbf{ 4}, (1999) 223.

\bibitem{Ono96:3449} K.\ Ono, H.\ Shimada, S.\ Kobayashi, and Y.\
Ootuka, J.\ Phys. Soc. Jpn. \textbf{ 65}, (1996) 3449; K.\ Ono, H.\
Shimada and Y.\ J.\ Ootuka, J.\ Phys.\ Soc.\ Jpn.\ \textbf{ 66}, 1261
(1997); L.\ F.\ Schelp, A.\ Fert, F.\ Fettar, P.\ Holody, S.\ F.\ Lee,
J.\ L.\ Maurice, F.\ Petroff and A.\ Vaures, Phys.\ Rev.\ B \textbf{
56}, (1997) R5747; S.\ Sankar, B.\ Dieny and A.\ E.\ Berkowitz, J.\
Appl.\ Phys.\ \textbf{ 81}, (1997) 5512.

\bibitem{Brataas99:93} A.\ Brataas, Yu.\ V.\ Nazarov, J.\ Inoue, and
G.\ E.\ W.\ Bauer, Phys.\ Rev.\ B \textbf{ 59}, (1999) 93; Eur.\ Phys.\
J.\ B \textbf{ 9}, (1999) 421; A.\ N.\ Korotkov and V.\ I.\ Safarov,
Phys.\ Rev.\ B \textbf{ 59}, (1999) 89; H.\ Imamura, S.\ Takahashi and S.\
Maekawa, Phys.\ Rev.\ B \textbf{ 59}, (1999) 6017; A.\ Brataas and X.\ H.\
Wang, cond-mat/0004082.

\bibitem{Barnas98:1058} J.\ Barnas and A.\ Fert, Phys.\ Rev.\ Lett.\
\textbf{ 80}, (1998) 1058; S.\ Takahashi and S.\ Maekawa, Phys.\ Rev.\
Lett.\ \textbf{ 80}, (1998) 1758; X.\ H.\ Wang and A.\ Brataas, Phys.\
Rev.\ Lett.\ \textbf{ 83}, (1999) 5138.

\bibitem{Johnson85:1790} M.\ Johnson and R.\ H.\ Silsbee, Phys.\ Rev.\
Lett.\ \textbf{ 55}, (1985) 1790; M.\ Johnson, Phys.\ Rev.\ Lett.\ \textbf{
70}, (1993) 2142; M.\ Johnson, Science \textbf{ 260}, (1993) 320.

\bibitem{Moodera96:4724} J.\ S.\ Moodera, L.\ R.\ Kinder, J.\ Appl.\
Phys.\ \textbf{ 79}, (1996) 4724.

\bibitem{Slon89:6995} J.\ C.\ Slonczewski, Phys. Rev. B \textbf{ 39},
(1989) 6995.

\bibitem{Valet93:7099} T.\ Valet and A.\ Fert, Phys.\ Rev.\ B \textbf{
48}, (1993) 7099; V.\ V.\ Ustinov and E.\ A.\ Kravtsov, J.\ Phys.\ :
Cond.\ Matter \textbf{ 7}, (1995) 3471; H.\ E.\ Camblong, P.\ M.\ Levy and
S.\ Zhang, Phys.\ Rev.\ B \textbf{ 51}, (1995) 16052.

\bibitem{Belzig99:1251} W. Belzig, F.K. Wilhelm, C. Bruder,
G. Sch\"{o}n, and A. Zaikin, Superl. Mircost. \textbf{ 25}, (1999) 1251.

\bibitem{Nazarov94:1420} Yu.\ V.\ Nazarov, Phys.\ Rev.\ Lett.\ \textbf{
73}, 1420 (1994); Superlattice Microst.\ \textbf{ 25}, (1999) 1221.

\bibitem{Huertas00:5700} D.\ Huertas Hernando, A.\ Brataas, Yu.\ V.\
Nazarov and G.\ E.\ W.\ Bauer, Phys.\  Rev.\ B \textbf{62}, (2000) 5700.

\bibitem{MacDonald9912391} A.\ H.\ MacDonald, cond-mat/9912392.

\bibitem{Brataas00:2481} A.\ Brataas, Yu.\ V. Nazarov, and G.\ E.\ W.\
Bauer, Phys.\ Rev.\ Lett.\ \textbf{ 84}, (2000) 2481.

\bibitem{Buttiker86:1761} M.\ B\"{u}ttiker, Phys.\ Rev.\ Lett.\
\textbf{ 57}, (1986) 1761.

\bibitem{Nazarov94:134} Yu.\ V.\ Nazarov, Phys.\ Rev.\ Lett.\ \textbf{
73}, (1994) 134.

\bibitem{Petrashov99:3281} V.\ T.\ Petrashov, I.\ A.\ Sosnin, I.\ Cox,
  A.\ Parsons, and C.\ Troadec, Phys.\ Rev.\ Lett.\ \textbf{ 83},
  (1999) 3281.

\bibitem{Bauer92:1676} G.\ E.\ W.\ Bauer, Phys.\ Rev.\ Lett.\ \textbf{
69}, (1992) 1676; M.\ J.\ M.\ de Jong and C.\ W.\ J.\ Beenakker,
Phys.\ Rev.\ Lett.\ \textbf{ 74}, (1995) 1657.

\bibitem{Brataas9906065} A.\ Brataas, Yu.\ V.\ Nazarov and Gerrit E.\
W.\ Bauer, cond-mat/9906065 (version 1).

\bibitem{Balents00:3463} L.\ Balents, R.\ Egger, Phys.\ Rev.\ Lett.\
\textbf{85}, (2000) 3463 .

\bibitem{Geux00:119} L.\ S.\ Geux, A.\ Brataas, and G.\ E.\ W.\ Bauer,
Acta.\ Phys.\ Pol.\ A \textbf{ 97}, (2000) 119.

\bibitem{Xia01} K.\ Xia, P.\ J.\ Kelly, G.\ E.\ W.\ Bauer, A.\
Brataas, and I.\ Turek (unpublished).

\bibitem{Waintal00:12317} X.\ Waintal, E.\ B.\ Myers, P.\ W.\ Brouwer,
and D.\ C.\ Ralph, Phys.\ Rev.\ B \textbf{62}, (2000) 12317.
 
\end{thebibliography}
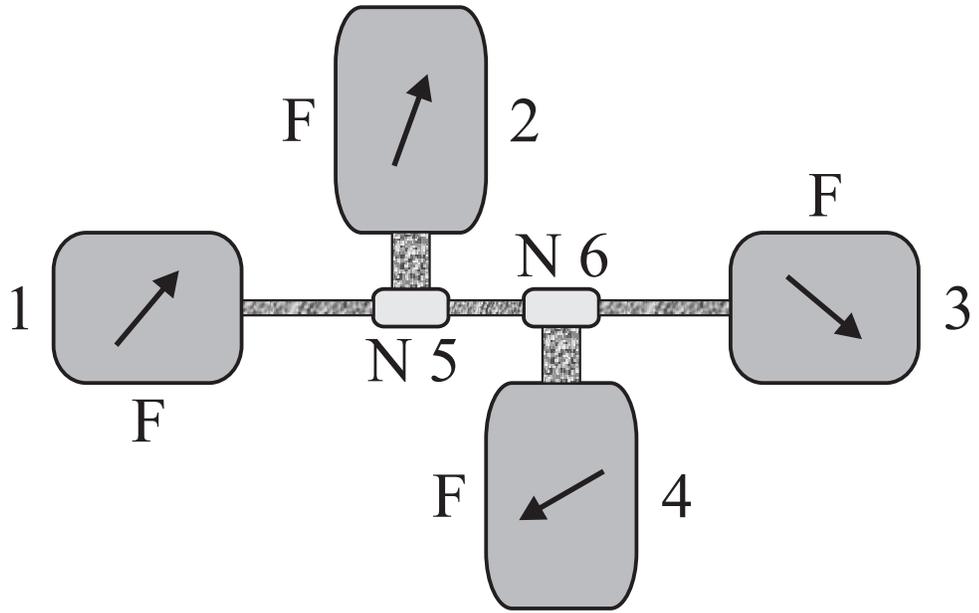
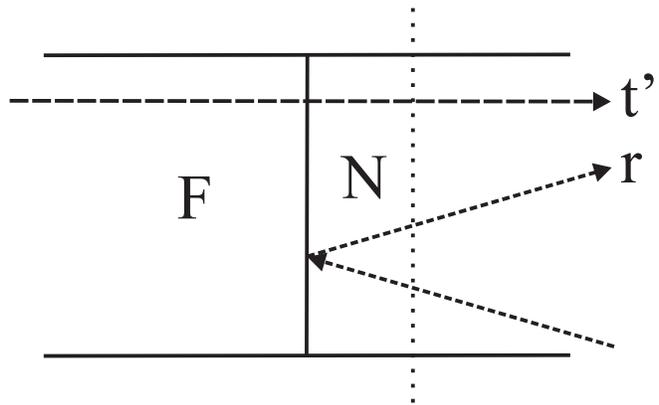
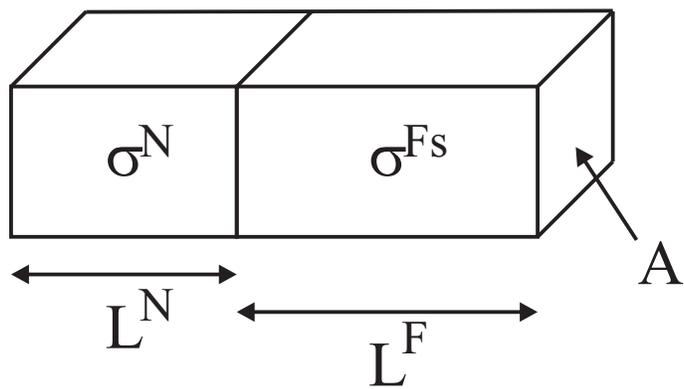
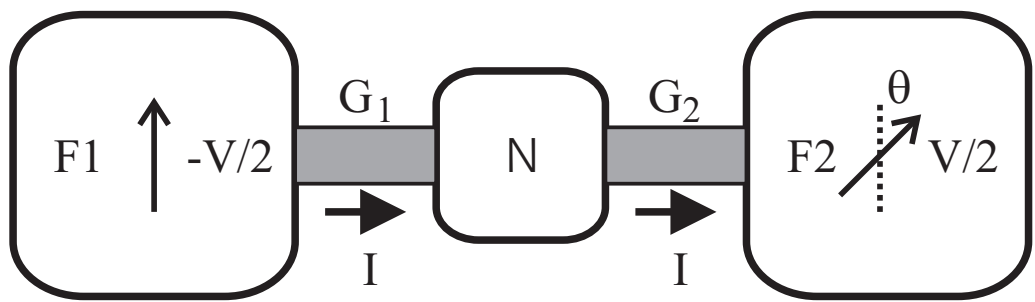
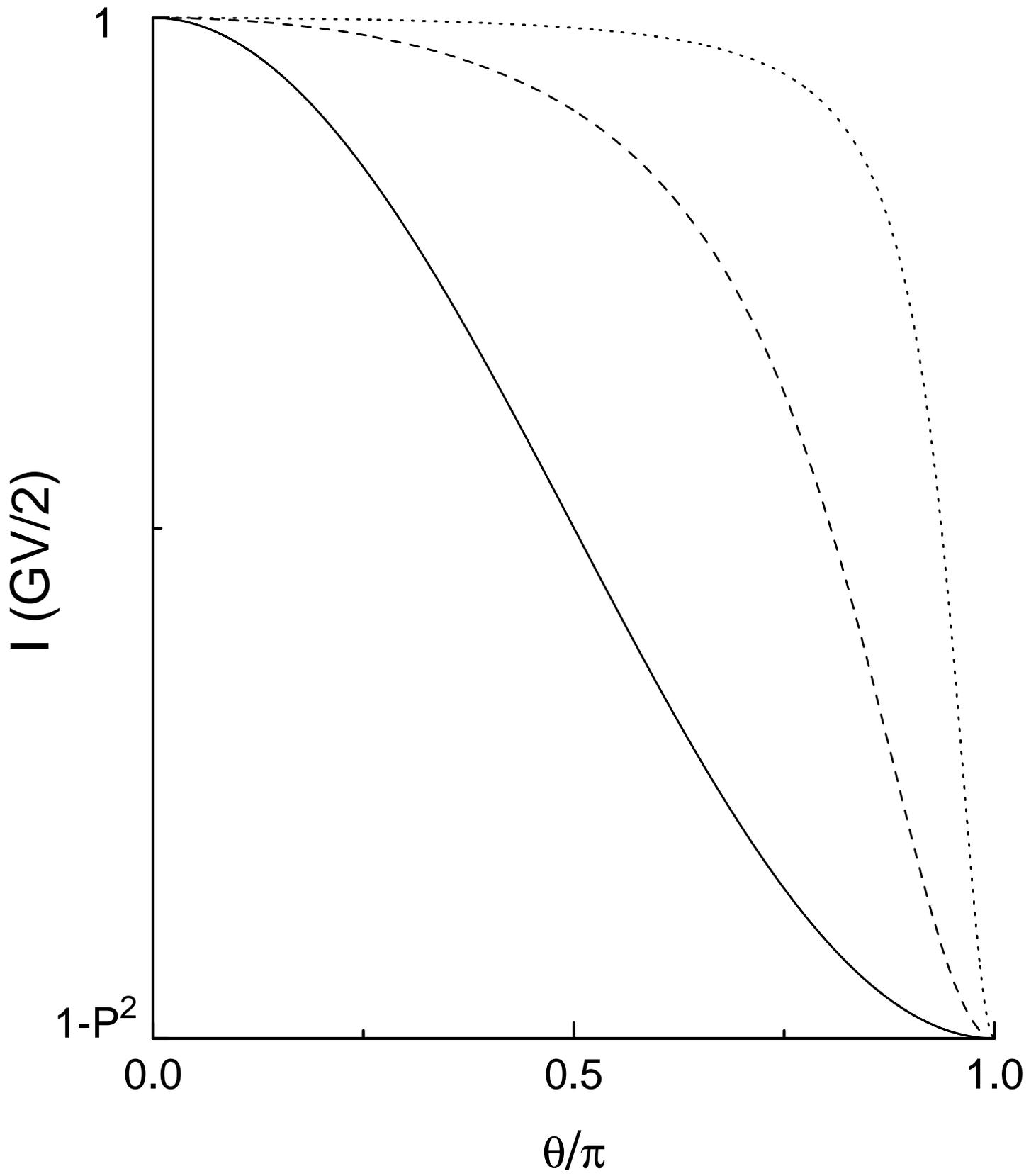
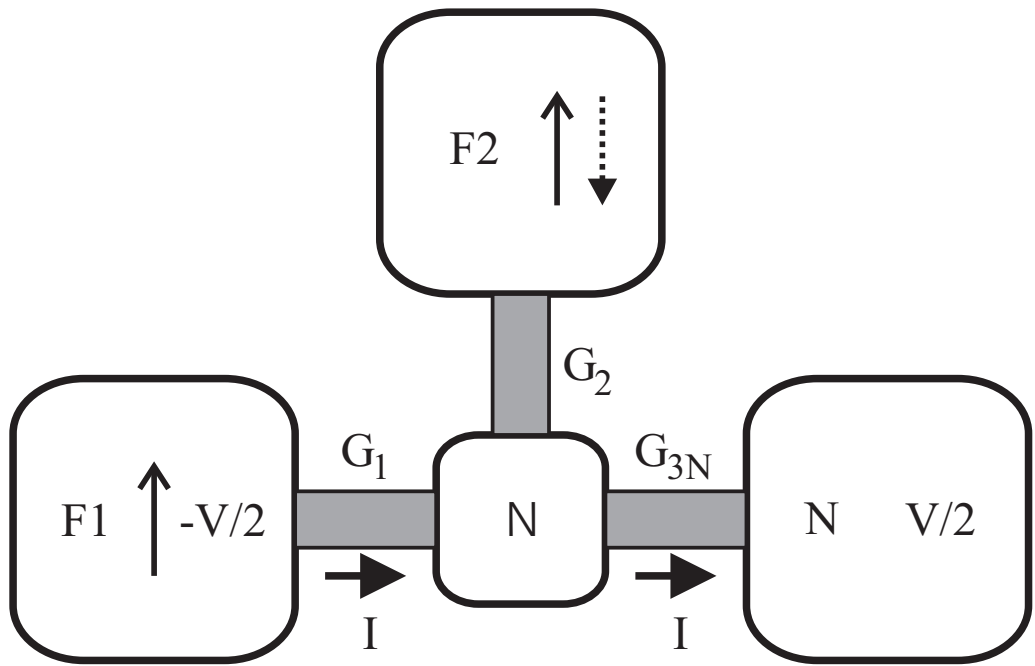
\end{document}